\documentclass[12pt]{article}
\usepackage{amsfonts}

\usepackage{amsmath}
\usepackage{graphicx}

\newtheorem{theorem}{Theorem}

\newtheorem{claim}[theorem]{Claim}

\newtheorem{definition}[theorem]{Definition}

\input{tcilatex}

\begin{document}

\baselineskip16pt

\begin{titlepage}

\  \\

\vspace{16 mm}

\begin{center}
{\Large \bf Corrections to the Abelian Born--Infeld Action}\\
\vspace{2mm}
{\Large \bf  Arising from Noncommutative Geometry}\\

\vspace{3mm}

\end{center}

\vspace{8 mm}

\begin{center}

Lorenzo Cornalba

\vspace{3mm}

{\small  Institut des Hautes Etudes Scientifiques} \\
{\small  cornalba@ihes.fr} \\

\vspace{5mm}

January 2000

\vspace{5mm}

\end{center}

\begin{abstract}

{ \it
In a recent paper Seiberg and Witten have argued that the full action 
describing the dynamics of coincident branes in the weak coupling 
regime is invariant under a specific
field redefinition, which replaces the group of ordinary gauge
transformations with 
the one of noncommutative gauge theory. This paper represents a first
step towards the
classification of invariant actions, in the simpler setting of the
abelian single
brane theory. In particular we consider a simplified model, in which the
group of 
noncommutative  gauge transformations is replaced with the group of 
symplectic diffeomorphisms of the brane world volume. We carefully define
what we mean, in this context, by invariant actions, and rederive the known
invariance of the Born--Infeld volume form. With the aid of a simple
algebraic tool, 
which is a generalization of the Poisson bracket on the brane world
volume, we 
are then able to describe invariant actions with an arbitrary number of derivatives.

}
\end{abstract}

\end{titlepage}

\section{Introduction\label{INTRO}}

The physics of branes with large background magnetic fields is intimately
connected, as shown in various works\footnote{%
For an extensive list of references, we refer the reader to \cite{SW}.} \cite
{SW,CDS}, to gauge theories on non--commutative spaces. In particular, it
has been shown, in a detailed study \cite{SW}, that there exists a large
freedom in the possible description of the physics of the gauge degrees of
freedom which live on the brane world--volume. One is free to choose the
non--commutativity parameter on the world--volume, and each possible choice
can be reached by a suitable gauge--orbit preserving field redefinition. The
most striking feature of the full action describing the brane dynamics at
small string coupling is that, regardless of the choice of the
non--commutativity parameter, it is, after the above--mentioned field
redefinition, invariant in form, in a sense which will be made sharper in
the later part of this introductory section. This property of the brane
action is highly non--trivial, and does constrain the action in a
considerable but not fully understood way. In particular it has been argued
in various settings \cite{SW,C} that, if one considers only terms without
derivatives, and if one looks at the $U(1)$ single brane theory, then the
unique action which is form invariant is the Born--Infeld one, which is
known to describe the low energy phenomena of brane physics.

It is of importance to understand how the invariance described in \cite{SW}
constrains the brane action in the more general non--abelian $U(N)$ context,
and also how it constrains higher derivative terms. There has been already
some results in this direction \cite{Okawa}, but the methods are not
systematic, and become of increasing complexity after the first few terms
have been constrained. This paper is a first step towards a classification
of invariant actions, in a simplified context in which the geometric nature
of the problem reduces the task to a manageable one. In particular we will
not address the non--abelian case, restricting ourselves to the $U(1)$
theory. Moreover we will work in a simplified setting, relying on a previous
note \cite{C,CS} by the author. In particular, we will substitute the group
of non--commutative gauge transformations with the simpler group of
symplectic diffeomorphisms of the brane world--volume, and we will carefully
describe what we mean by invariant actions in this case. With the aid of a
simple algebraic tool, which is a generalization of the natural Poisson
bracket on the brane world--volume, we will then be able to generate in a
simple and powerful way invariant actions with an arbitrary number of
derivatives.

Let us note that the classification of invariant actions is intimately tied
to a deeper understanding of T--duality in the context of open--string
physics \cite{C2}. This can be better understood if we toroidally compactify
space--time. It is then true that, for some integral values of the
magnetic $%
B$ field, one can consider the brane configuration as a bound state of
higher dimensional branes with branes of lower dimension. T--duality then
exchanges the two types of branes, and therefore also changes the underlying
gauge group. There must therefore exist a (highly non--local) field
redefinition which maps gauge orbits of one gauge group to gauge orbits of
the other gauge group. Moreover the form of the action must be invariant
under T--duality, and therefore the field--redefinition must respect the
form of the action. Again, this is a highly non--trivial requirement,
and it
can be shown \cite{C2} to be equivalent, using simple Morita equivalence
arguments, to the statements described in \cite{SW}.

This paper has the following structure. We conclude this section by
recalling the results of \cite{SW}, both to set notation and to clarify what
we mean by invariance of the brane action. In section \ref{SETTING} we then
briefly recall the work \cite{C} and describe the simplified setting within
which we shall consider the problem. The invariance of the Born--Infeld
action is then shown in section \ref{BI}, and it is used to give a clear
definition, in section \ref{INVARIANCE}, of what we mean by invariant
actions within the setting of this paper. In section \ref{BRACKET} we
finally introduce the generalized bracket and we show how it can be used to
construct invariant brane actions with an arbitrary number of
derivatives. A
few examples with two derivatives are then considered in section \ref
{EXAMPLES}. Conclusions and open problems are left for the final section \ref
{CONCLUSION}.

Let us then proceed to a quick review of \cite{SW}. We will work throughout
with units such that 
\begin{equation*}
2\pi \alpha ^{\prime }=1.
\end{equation*}
Let $M=\mathbb{R}^{n}$ be the flat space--time manifold, parametrized by
coordinates $x^{a}$, and with constant background metric and NS two--form
given by the matrices $g_{ab}$ and $B_{ab}$ (we will assume that
$B_{ab}$ is
invertible). The arguments that follow do not rely on supersymmetry
considerations, and are valid both in the context of bosonic string theory
as well as in the context of Type II\ superstring theories. We then indicate
with $n$ the space--time dimension, with the understanding that $n=26$
or $%
n=10$.

We shall not be interested in the physics of the closed string sector, and
we will accordingly leave the geometry of the background space--time
manifold fixed. We will, on the other hand, concentrate on the dynamics of
open string sector of the theory, by introducing $N$ branes of maximal size
-- \textit{i.e. }such that the brane world--volume coincides with the
space--time manifold $M$. The dynamical degrees of freedom are then
described by a $U(N)$ connection on $M$. In the weak coupling regime $%
g_{s}\rightarrow 0$ the interaction of the brane gauge bosons are computed
from string theory disk diagrams, and can be reconstructed from a low energy
effective action of the general form 
\begin{equation}
S=\frac{1}{g_{s}}\int d^{n}x\sqrt{\det g_{ab}}\text{Tr}\left(
1+c^{abcd}\omega _{ab}\omega _{cd}+\cdots \right) ,  \label{eq300}
\end{equation}
where 
\begin{equation*}
\omega =F+B
\end{equation*}
and the coefficients $c^{abcd},\cdots $ are constructed from the tensor $%
g_{ab}$ (for example the first coefficient is $\frac{1}{4}g^{ac}g^{bd}$) \
As indicated by the notation, the complete effect of the NS two--form $B\,$%
is obtained by replacing the $U(N)$ field strength $F$ with $\omega
=F+B$ in
the action.

The above action is defined only up to field redefinitions. The simplest
type of redefinition, which had been already considered extensively in the
works \cite{T,AT1,AT2}, are gauge covariant and leave the general form of
the action invariant, with the unique effect of changing some of the
coefficients. The redefinition is of the form $A_{a}\rightarrow
A_{a}+d{}^{bc}D_{b}F_{ac}+\cdots $, where again the coefficients $%
d^{bc},\cdots $ are constructed in terms of the metric. A more powerful
possible field redefinition has been shown to exist in the recent work \cite
{SW}. The change of variables does not preserve the group of gauge
transformations, but on the other hand it substitutes it with the group of
gauge transformations of non--commutative gauge theory on the
world--volume $%
M$. More precisely, there is a transformation $A_{a}\rightarrow \widehat{A}%
_{a}$ (which we shall call Seiberg--Witten transformation) preserving gauge
orbits such that, in terms of the non--commutative field strength $\widehat{F%
}$, or, better, of the combination 
\begin{equation*}
\Omega =\widehat{F}-B,
\end{equation*}
the action reads 
\begin{equation*}
S=\frac{1}{G_{s}}\int d^{n}\sigma \sqrt{\det G_{ab}}\text{Tr}\left(
1+C^{abcd}\Omega _{ab}\star \Omega _{cd}+\cdots \right) .
\end{equation*}
In the above, the new metric tensor $G_{ab}$ and string coupling
constant $%
G_{s}$ are given by 
\begin{eqnarray}
G &=&-B\frac{1}{g}B  \notag \\
\frac{1}{g_{s}}\sqrt{\det B} &=&\frac{1}{G_{s}}\sqrt{\det G}.  \label{eq100}
\end{eqnarray}
Moreover, the coefficients $C^{abcd}$ are obtained starting from the
coefficients $c^{abcd}$ and replacing the metric $g_{ab}$ with $G_{ab}$.
Finally, the non--commutativity parameter defining the product $\star $ and
the field--strength $\widehat{F}$ is given by\footnote{%
We recall that $f\star g=\exp (\frac{i}{2}\theta ^{ij}\partial
_{i}^{f}\partial _{j}^{g})f\cdot g$ and that $\widehat{F}_{ab}=\partial _{a}%
\widehat{A}_{b}-\partial _{b}\widehat{A}_{a}-i\widehat{A}_{a}\star \widehat{A%
}_{b}+i\widehat{A}_{b}\star \widehat{A}_{a}$.} 
\begin{equation*}
\theta =\frac{1}{B}.
\end{equation*}
In the work \cite{SW} the transformation $A_{a}\rightarrow \widehat{A}_{a}$
is determined using the two requirements that it must preserve gauge orbits
and that it must be expressible as a power series in $\theta $, with the
coefficients of the series being local expressions in the fields. These
requirements clearly defines the map up to gauge--covariant local field
redefinitions. On the other hand a more precise statement of \cite{SW} says
that there is, among the possible maps $A_{a}\rightarrow \widehat{A}_{a}$,
one for which the action is form--invariant, in the sense described above.

The problem of invariance has not been analyzed in any detail. We will now
describe, in the next sections, a simplified setting which will allow us to
tackle the problem in a simple but powerful way.

\section{The Simplified Setting\label{SETTING}}

We will, throughout the rest of this paper, work in the simplified context
of the abelian $U(1)$ theory. Although this choice does imply a considerable
loss of information, we will see that the abelian theory has already a rich
structure, and does provide partial information about the non--abelian case.

The second simplification concerns the map $A_{a}\rightarrow \widehat{A}_{a}$%
, and follows the author's previous note \cite{C}. In this section we
quickly review the results of \cite{C}, and we rephrase them in the context
of the problem at hand. The starting point of \cite{C} is the observation
that the two--form $\omega =B+F$ defines a symplectic structure on $M$. On
one hand $\omega $ is clearly closed. Moreover, since we always work
perturbatively in $F$, and since $B$ is invertible, one can take the formal
inverse of $\omega $, thus showing that $\omega $ is non--degenerate.
Therefore, by Darboux's theorem, one can find coordinates $\sigma ^{i}$
on $%
M $ such that\footnote{%
We will use the following general conventions concerning coordinate systems
and indices. A general coordinate system on $M$ will be denoted by $\xi
^{\alpha }$, and in general will have Greek indices $\alpha ,\beta
,\cdots $%
. The fixed coordinate system $x^{a}$ will be called flat, and will have in
general roman indices $a,b,\cdots $. Finally coordinate systems $\sigma ^{i}$
for which the two--form $\omega $ has constant coefficients $B_{ij}$
will be
called symplectic, and will have roman indices $i,j,\cdots $.} 
\begin{equation*}
\omega =\frac{1}{2}B_{ij}\,d\sigma ^{i}\wedge d\sigma ^{j}.
\end{equation*}
In these new coordinates, the fluctuations of the field strength $F$ have
been replaced by the parallel displacements of the brane, which are
described by the coordinate functions $x^{a}(\sigma )$. Moreover the
coordinates $\sigma ^{i}$ are clearly defined up to symplectic
diffeomorphisms of $M$. The original group of abelian gauge transformations
is replaced by the group of symplectomorphisms of $(M,\omega )$, and the
correspondence between $A_{a}(x)$ and $x^{a}(\sigma )$ respects the gauge
orbits of the two group actions. One is therefore in a situation similar to
the one considered in \cite{SW}, with the simplifying difference that the
group of non--commutative gauge transformations is replaced by the group of
symplectomorphisms of the brane world--volume.

To make contact with the notation of the previous section one defines the
Poisson bracket\thinspace $\{,\}$ with respect to the symplectic structure
on $M$%
\begin{equation*}
\{f,g\}=\left( \frac{1}{\omega }\right) ^{\alpha \beta }\partial _{\alpha
}f\,\partial _{\beta }g.
\end{equation*}
The above formula is particularly simple if one uses the symplectic
$\sigma $%
--coordinates, and it then reads 
\begin{equation*}
\{f,g\}=\theta ^{ij}\partial _{i}f\,\partial _{j}g.
\end{equation*}
If one defines the non--commutative gauge potential $\widehat{A}_{a}$ by 
\begin{equation*}
x^{a}(\sigma )=\sigma ^{a}+\theta ^{ab}\widehat{A}_{b}(\sigma )
\end{equation*}
and the corresponding field strength by\footnote{%
This is clearly an exception to the index convention, which is forced by the
notation.} 
\begin{equation*}
\widehat{F}_{ab}=\partial _{a}\widehat{A}_{b}-\partial _{b}\widehat{A}%
_{a}+\left\{ \widehat{A}_{a},\widehat{A}_{b}\right\} ,
\end{equation*}
one finds that 
\begin{equation*}
\{x_{a},x_{b}\}=\Omega _{ab}=\widehat{F}_{ab}-B_{ab},
\end{equation*}
where we have lowered the index on the coordinate function $x^{a}$ using $%
B_{ab}$ 
\begin{equation*}
x_{a}=B_{ab}x^{b}.
\end{equation*}

We have quickly reviewed the results of \cite{C} and we are therefore in a
position, given the above notation, to rephrase the meaning of the
invariance of the action given in section \ref{INTRO} in this new framework,
starting from the simple invariance of the Born--Infeld volume form.

\section{Invariance of the Born--Infeld Volume Form\label{BI}}

In this section we prove the exact invariance of the Born--Infeld volume
form under the change of coordinates described in the previous section.
Before we do so, let us though clarify one point of notation.

In order to limit the number of symbols, we will use, as a general rule, the
same letter to indicate an abstract tensor, and its components in a specific
coordinate system, and we will rely on our index convention to distinguish
among coordinate systems. In some cases though this might be confusing,
given the standard notation in the subject. For example the metric
tensor $%
g=g_{ab}dx^{a}\otimes dx^{b}$ reads, in a symplectic coordinate system
(recall $G_{ab}=B_{ac}B_{bd}g^{cd}$) 
\begin{equation*}
g=g_{ab}\partial _{i}x^{a}\partial _{j}x^{b}\,d\sigma ^{i}\otimes d\sigma
^{j}=G^{ab}\partial _{i}x_{a}\partial _{j}x_{b}\,d\sigma ^{i}\otimes d\sigma
^{j}.
\end{equation*}
Following the general rule, we could\ use the symbol $g_{ij}$ for $%
G^{ab}\partial _{i}x_{a}\partial _{j}x_{b}$. We will not do this, and we
will reserve the letter $g_{ab}$ for the constant metric in the flat
coordinate system, and will always write $G^{ab}\partial _{i}x_{a}\partial
_{j}x_{b}$ when using $\sigma $--coordinates. Similarly we will use $B_{ij}$
instead of $\omega _{ij}$.

With this in mind, let us consider the Born--Infeld volume form 
\begin{equation*}
\Phi =\frac{1}{g_{s}}d^{n}\xi \sqrt{\det
(g+B+F)}=\frac{1}{g_{s}}d^{n}\xi 
\sqrt{\det (g+\omega )}.
\end{equation*}
In flat coordinates $x^{a}$ one has 
\begin{equation*}
\Phi =\frac{1}{g_{s}}d^{n}x\sqrt{\det (g+\omega )_{ab}}.
\end{equation*}
Let us now compute the volume form $\Phi $ in the symplectic coordinates $%
\sigma ^{i}$. If we introduce the matrix--valued field 
\begin{equation*}
M^{i}{}_{j}=\theta ^{ik}G^{ab}\partial _{k}x_{a}\partial _{j}x_{b}
\end{equation*}
we easily compute that 
\begin{eqnarray*}
\text{Tr}\,M &=&G^{ab}\Omega _{ba}=\text{Tr}\frac{1}{G}\Omega \\
\text{Tr}\,M^{2} &=&G^{ab}G^{cd}\text{Tr}\,\theta \partial x_{a}\partial
x_{b}\theta \partial x_{c}\partial x_{d}=G^{ab}\Omega _{bc}G^{cd}\Omega
_{da}=\text{Tr}\frac{1}{G}\Omega \frac{1}{G}\Omega \\
\text{Tr}\,M^{n} &=&\text{Tr}\left( \frac{1}{G}\Omega \right) ^{n}.
\end{eqnarray*}
Using the above equation one can show that 
\begin{equation*}
\sqrt{\det (1+\theta G^{ab}\partial x_{a}\partial x_{b})}=\sqrt{\det \left(
1+\frac{1}{G}\Omega \right) }
\end{equation*}
by expanding $\det {}^{1/2}(1+A)$ in terms of traces of powers of $A$. We
may then check that, in the symplectic coordinates $\sigma ^{i}$, the
Born--Infeld form reads (using equation \ref{eq100}) 
\begin{eqnarray*}
\Phi &=&\frac{1}{g_{s}}d^{n}\sigma \sqrt{\det (B+G^{ab}\partial
x_{a}\partial x_{b})_{{}}} \\
&=&\frac{1}{G_{s}}d^{n}\sigma \sqrt{\det G(1+\theta G^{ab}\partial
x_{a}\partial x_{b})} \\
&=&\frac{1}{G_{s}}d^{n}\sigma \sqrt{\det G\left( 1+\frac{1}{G}\Omega
\right) 
} \\
&=&\frac{1}{G_{s}}d^{n}\sigma \sqrt{\det (G+\Omega )},
\end{eqnarray*}
thus proving the invariance of the Born--Infeld action under the simplified
Seiberg--Witten change of variables described in the previous section.

\section{Statement of the Invariance Problem\label{INVARIANCE}}

We now have the notation needed to define the concept of invariant action
within the setting of this paper, and to describe the problem that we wish
to address. We will not give general definitions and proofs, but we will
work with some meaningful examples, with the hope that the general case can
be easily understood from them.

We start with a basic observation, by noting that 
\begin{equation*}
\{\sigma ^{i},f\}=\theta ^{ij}\partial _{j}f,
\end{equation*}
and therefore that 
\begin{equation*}
\{x_{a},f\}=\partial _{a}f+\{\widehat{A}_{a},f\}\equiv \widehat{D}_{a}f.
\end{equation*}
Then, in the limit $B\rightarrow \infty $, $\theta \rightarrow 0$, one has
that\footnote{%
The correspondence $\{x_{a},=\widehat{D}_{a}\rightarrow \partial _{a}$
is a
notable exception to the index convention, since we are using an index $%
a,b,\cdots $ in a symplectic coordinate system. This is the same exception
noted in the previous footnote, since $\widehat{F}_{ab}=\widehat{D}_{a}%
\widehat{D}_{b}-\widehat{D}_{b}\widehat{D}_{a}$.} 
\begin{equation*}
\{x_{a},f\}\rightarrow \partial _{a}f.
\end{equation*}
This means that the correct dual for a generic derivative term 
\begin{equation*}
\partial _{a}\partial _{b}\omega _{cd}
\end{equation*}
is given by 
\begin{equation*}
\{x_{a},\{x_{b},\Omega _{cd}\}\}=\{x_{a},\{x_{b},\{x_{c},x_{d}\}\}\}.
\end{equation*}
At the level of the action $S$ describing (equation \ref{eq100})the brane
world--volume degrees of freedom we may consider a generic term 
\begin{equation*}
\frac{1}{g_{s}}\int d^{n}x\sqrt{\det g_{ab}}g^{ad}g^{be}g^{cf}\partial
_{a}\omega _{bc}\partial _{d}\omega _{ef}.
\end{equation*}
Following the above observations, this term in the action should correspond,
in the symplectic coordinates $\sigma ^{i}$, to the term 
\begin{equation*}
\frac{1}{G_{s}}\int d^{n}\sigma \sqrt{\det G_{ab}}G^{ad}G^{be}G^{cf}\{x_{a},%
\Omega _{bc}\}\{x_{d},\Omega _{ef}\}.
\end{equation*}
In the above discussion we have not addressed an important issue, which
depends on the fact that we are not considering the full non--abelian
theory, but that we are uniquely concentrating on the abelian $U(1)$ theory,
and that we are therefore neglecting commutator terms. More precisely, let
us note, for example, that the correspondence 
\begin{equation*}
\partial _{a}\partial _{b}\omega _{cd}\rightarrow \{x_{a},\{x_{b},\Omega
_{cd}\}\}
\end{equation*}
is not well--defined. In fact, although the partial derivatives $\partial
_{a}\partial _{b}$ commute, the corresponding Poisson bracket
derivatives $%
\{x_{a},\{x_{b},\cdots \}\}$ do not. On the other hand, the commutator is
proportional, by the Jacobi identity, to 
\begin{equation*}
\{\Omega _{ab},\cdots \}
\end{equation*}
and therefore vanishes in the limit $B\rightarrow \infty $, $\theta
\rightarrow 0$. We can therefore state a precise form of the invariance
concept which we wish to analyze.

\begin{definition}
The abelian action $S$ for a single brane is invariant under the simplified
Seiberg--Witten transformation described in section \ref{SETTING} if it has,
when written in terms of the symplectic coordinates $\sigma ^{i}$, the same
form (as described in this section) as in the original flat coordinates $%
x^{a}$, up to terms which vanish in the $B\rightarrow \infty $, $\theta
\rightarrow 0$ limit.
\end{definition}

In the next section we will see that, with the aid of a simple algebraic
tool, we will be able to easily generate actions which do possess the
property just described.

\section{The Generalized Bracket and Invariant Actions\label{BRACKET}}

In order to construct in a systematic way actions which are invariant in the
sense described above, we introduce a bilinear differential operation
defined on functions, which generalizes the Poisson bracket $\{,\}$. Given
two functions $f$ and $g$ on $M$ we define the bracket $[f,g]$ by 
\begin{equation*}
\left[ f,g\right] =\left( \frac{1}{g+\omega }\right) ^{\alpha \beta
}\partial _{\alpha }f\,\partial _{\beta }g.
\end{equation*}
Let me note that the above bilinear form is not antisymmetric 
\begin{equation*}
\left[ f,g\right] \neq -\left[ g,f\right]
\end{equation*}
and does not satisfy the Jacobi identity. On the other hand it will be an
extremely useful tool in constructing invariant actions.

An intuitive argument for the above definition is the following. We know
that the effect of the NS two--form $B$ is described with the
replacement $%
F\rightarrow \omega =F+B$. This is justified by looking at the string
conformal field theory $\int_{\Sigma }B+\int_{\partial \Sigma }A$ \
($\Sigma 
$ is the string world--volume) and by noting that the transformation $%
B\rightarrow B+d\Lambda $, $\ A\rightarrow A-\Lambda $, leaves both the
action and $\omega $ invariant. On the other hand, if one considers the
massless closed string sector vertex operators $(h_{ab}+B_{ab})\int
d^{2}z\,\partial X^{a}\overline{\partial }X^{b}e^{ik\cdot X}$ one notices
that the natural combination\footnote{%
The combination $E=g+B$ is also relevant in T--duality on tori, where it
transforms as $E\rightarrow E^{-1}$.} which appears is $g+B$. Therefore, one
is lead to look at expression $g+B+F=g+\omega $, which is present both in
the Born--Infeld volume form, and in the bracket $[,]$ which we have just
described.

Let us start with a first basic example, by computing the bracket $%
[x^{a},x^{b}]$. In the flat coordinates $x^{a}$ one simply obtains 
\begin{equation*}
\left[ x^{a},x^{b}\right] =\left( \frac{1}{g+\omega }\right) ^{ab}.
\end{equation*}
On the other hand one can compute the bracket $[,]$ in the symplectic
coordinates $\sigma ^{i}$. In this case it will be easier to consider the
quantity $[x_{a},x_{b}]$, where we have lowered the indices with $%
x_{a}=B_{ab}x^{b}$. From the definition we obtain that 
\begin{equation*}
\left[ x_{a},x_{b}\right] =\left( \frac{1}{\partial x_{c}\partial
x_{d}G^{cd}+B}\right) ^{ij}\partial _{i}x_{a}\,\partial _{j}x_{b}.
\end{equation*}
We can then expand the denominator in powers of the induced metric $\partial
x_{c}\partial x_{d}G^{cd}$ and obtain 
\begin{eqnarray*}
\left[ x_{a},x_{b}\right] &=&\left( \theta -\theta \partial x_{c}\partial
x_{d}G^{cd}\theta +\cdots \right) ^{ij}\partial _{i}x_{a}\,\partial _{j}x_{b}
\\
&=&\{x_{a},x_{b}\}-\{x_{a},x_{c}\}G^{cd}\{x_{d},x_{b}\}+\cdots \\
&=&\left( \Omega -\Omega \frac{1}{G}\Omega +\cdots \right) _{ab} \\
&=&\left[ G\left( \frac{1}{G}-\frac{1}{G+\Omega }\right) G\right] _{ab}.
\end{eqnarray*}
For reasons which we will shortly describe, we define the functions $y^{a}$
by 
\begin{equation*}
y^{a}=G^{ab}x_{b}=G^{ab}B_{bc}x^{c}.
\end{equation*}
We can then use the previous computation and write 
\begin{equation}
\lbrack y^{a},y^{b}]=\left( \frac{1}{G}-\frac{1}{G+\Omega }\right) ^{ab}.
\label{eq200}
\end{equation}

The importance of the functions $y^{a}$ is that they play, in a symplectic
coordinate system, the same role played by the coordinates $x^{a}$ in a flat
coordinate system. More precisely, all contractions of the functions $x^{a}$
with metric tensors $g_{ab}$ are equal to equivalent contractions of the
functions $y^{a}$ with the tensor $G_{ab}=B_{ac}B_{bd}g^{cd}$. For
example 
\begin{equation*}
g_{ab}\left[ x^{a},x^{b}\right] =G^{ab}[x_{a},x_{b}]=G_{ab}[y^{a},y^{b}].
\end{equation*}
We can then consider the simple action 
\begin{equation*}
\int \Phi \,g_{ab}\left[ x^{a},x^{b}\right] =\int \Phi \,G_{ab}[y^{a},y^{b}].
\end{equation*}
We have proved in the last section the invariance of the Born--Infeld volume
form. If one neglects the term $\left( \frac{1}{G}\right) ^{ab}$ and the
minus sign in the expression \ref{eq200} for $[y^{a},y^{b}]$, the previous
computation would shows that the above action is invariant. Clearly though
one cannot neglect in general the constant $\left( \frac{1}{G}\right) ^{ab}$%
. The correct solution is then to consider derivative terms, which
automatically eliminate the constant part in the expression for the
bracket $%
[y^{a},y^{b}]$. In the framework of this paper this fact is yet an other
indication that the Born--Infeld action is the unique action without
derivatives which is invariant under the Seiberg--Witten transformations.

Let us now move to the analysis of derivative terms. To this end we consider
a second example, and we compute the double bracket 
\begin{equation*}
\left[ x^{a},\left[ x^{b},x^{c}\right] \right] .
\end{equation*}
In the flat coordinate system one simply has 
\begin{equation*}
\left[ x^{a},\left[ x^{b},x^{c}\right] \right] =\left( \frac{1}{g+\omega }%
\right) ^{ad}\partial _{d}\left( \frac{1}{g+\omega }\right) ^{bc}.
\end{equation*}
Using the previous results, and computing in the symplectic coordinate
system, one has, on the other hand, 
\begin{eqnarray*}
\left[ x_{a},\left[ y^{b},y^{c}\right] \right] &=&-\left( \frac{1}{\partial
x_{f}\partial x_{d}G^{fd}+B}\right) ^{ij}\partial _{i}x_{a}\partial
_{j}\left( \frac{1}{G+\Omega }\right) ^{bc} \\
&=&-\{x_{a},\left( \frac{1}{G+\Omega }\right)
^{bc}\}+\{x_{a},x_{f}\}G^{fd}\{x_{d},\left( \frac{1}{G+\Omega }\right)
^{bc}\}+\cdots .
\end{eqnarray*}
Finally, rewriting the result solely in terms of the coordinates $y^{a}$,
one obtains 
\begin{eqnarray*}
\left[ y^{a},\left[ y^{b},y^{c}\right] \right] &=&\left(
-G^{ad}+G^{ae}\{x_{e},x_{f}\}G^{fd}+\cdots \right) \{x_{d},\left( \frac{1}{%
G+\Omega }\right) ^{bc}\} \\
&=&-\left( \frac{1}{G+\Omega }\right) ^{ad}\{x_{d},\left(
\frac{1}{G+\Omega }%
\right) ^{bc}\}.
\end{eqnarray*}
We now see that, recalling the correspondence 
\begin{equation*}
\{x_{a},\,\rightarrow \partial _{a}
\end{equation*}
one has invariance of the term $\left[ x^{a},\left[ x^{b},x^{c}\right] %
\right] $, up to a minus sign.

Let us now move to a third and last example, by computing the triple
bracket 
\begin{equation*}
\left[ \left[ x^{a},x^{b}\right] ,\left[ x^{c},x^{d}\right] \right] .
\end{equation*}
As before, in flat coordinates, the expression is easily computed as 
\begin{equation*}
\left[ \left[ x^{a},x^{b}\right] ,\left[ x^{c},x^{d}\right] \right]
=\left( 
\frac{1}{g+\omega }\right) ^{ef}\partial _{e}\left( \frac{1}{g+\omega }%
\right) ^{ab}\partial _{f}\left( \frac{1}{g+\omega }\right) ^{cd}.
\end{equation*}
Using again the previous results, and following the usual procedure of
expanding the generalized bracket in powers of the induced metric, one
obtains 
\begin{eqnarray*}
\left[ \left[ y^{a},y^{b}\right] ,\left[ y^{c},y^{d}\right] \right]
&=&\left( \frac{1}{\partial x_{g}\partial x_{h}G^{gh}+B}\right)
^{ij}\partial _{i}\left( \frac{1}{G+\Omega }\right) ^{ab}\partial
_{j}\left( 
\frac{1}{G+\Omega }\right) ^{cd} \\
&=&\left\{ \left( \frac{1}{G+\Omega }\right) ^{ab},\left(
\frac{1}{G+\Omega }%
\right) ^{cd}\right\} + \\
&&+\left( \frac{1}{G+\Omega }\right) ^{ef}\left\{ x_{e},\left( \frac{1}{%
G+\Omega }\right) ^{ab}\right\} \left\{ x_{f},\left( \frac{1}{G+\Omega }%
\right) ^{cd}\right\}
\end{eqnarray*}
In this case we have complete invariance of the term under consideration.
Let us note that a new feature of this last example is that the expression
in terms of the variables $y^{a}$ contains an explicit commutator term $%
\left\{ \left( \frac{1}{G+\Omega }\right) ^{ab},\left( \frac{1}{G+\Omega }%
\right) ^{cd}\right\} $, which is not present in the same expression in
terms of the variables $x^{a}$. On the other hand the commutator
vanishes in
the $B\rightarrow \infty $, $\theta \rightarrow 0$ limit, and therefore
should be neglected, as already noted in the definition given in the last
section. We are therefore ready to state the following

\begin{claim}
Consider a product $\Pi $ of terms, each of which is a multiple bracket of
the coordinate functions $x^{a}$ (for example one term could be of the
form $%
\left[ \left[ x^{a},x^{b}\right] ,\left[ x^{c},x^{d}\right] \right] $ or $%
[x^{a},[x^{b},x^{c}]]$). Moreover, let all the free indices $a,b,\cdots
$ be
contracted using the metric tensor $g_{ab}$. Let us further assume that

\begin{enumerate}
\item  No single term in $\Pi $ is a single bracket of the form $%
[x^{a},x^{b}]$ (all terms are derivative terms, so that the constant
part in 
$[y^{a},y^{b}]$ does not spoil invariance).

\item  There is an even number of basic brackets $[x^{a},x^{b}]$ (which are,
by $1$, necessarily contained within other brackets).

Then the action 
\begin{equation*}
\int \Phi \,\ \Pi
\end{equation*}

is invariant under the simplified Seiberg--Witten transformations.
\end{enumerate}
\end{claim}

Let us note that, given two actions $S_{1}$ and $S_{2}$ which are invariant,
so is the linear combinations $aS_{1}+bS_{2}$. Therefore we should consider
in general actions of the form $\int \Phi \left( a_{1}\Pi _{1}+a_{2}\Pi
_{2}+\cdots \right) $, where the coefficients $a_{i}$ must be determined
from other considerations.

\section{Some Examples\label{EXAMPLES}}

In this last section we analyze in some detail two examples of simple
invariant actions with two derivatives, in order to make contact with
actions of the form \ref{eq300}. We will work only in the flat coordinate
system, and we will therefore neglect the index convention, using both
indices $a,b,\cdots $ and indices $i,j,\cdots $.

Let us start by considering the action 
\begin{equation*}
\int \Phi \,\ g_{ac}g_{bd}\left[ \left[ x^{a},x^{b}\right] ,\left[
x^{c},x^{d}\right] \right]
\end{equation*}
We will for simplicity of notation, but with no loss in generality, take $%
g_{s}=1$ and $g_{ab}=\delta _{ab}$. Moreover we will write all the equations
in the case $B_{ab}=0$, therefore replacing $\omega $ with $F$. The
lagrangian then becomes 
\begin{equation*}
\mathcal{L}_{1}=\sqrt{\det (g+F)}\,g_{ac}g_{bd}\left( \frac{1}{g+F}\right)
^{ij}\partial _{i}\left( \frac{1}{g+F}\right) ^{ab}\partial _{j}\left( \frac{%
1}{g+F}\right) ^{cd}.
\end{equation*}
The following quick computation 
\begin{eqnarray*}
g_{ac}g_{bd}\partial _{i}\left( \frac{1}{g+F}\right) ^{ab}\partial
_{j}\left( \frac{1}{g+F}\right) ^{cd} &\simeq &\partial _{i}\left(
F-FF\right) _{ab}\partial _{j}\left( F-FF\right) _{ab} \\
&=&\partial _{i}F_{ab}\partial _{j}F_{ab}+\partial _{i}\left(
F_{ac}F_{cb}\right) \partial _{j}\left( F_{ad}F_{db}\right)
\end{eqnarray*}
can then be used to show that 
\begin{eqnarray}
\mathcal{L}_{1} &=&\partial _{i}F_{ab}\partial _{i}F_{ab}+\frac{1}{4}%
F_{cd}F_{cd}\partial _{i}F_{ab}\partial _{i}F_{ab}+F_{ik}F_{kj}\partial
_{i}F_{ab}\partial _{j}F_{ab}+  \label{eq400} \\
&&+2F_{ab}F_{bc}\partial _{i}F_{cd}\partial _{i}F_{da}+2F_{bc}F_{da}\partial
_{i}F_{ab}\partial _{i}F_{cd}+\cdots ,  \notag
\end{eqnarray}
where $\cdots $ denotes terms of order $F^{2n}\partial F\partial F$ with $%
n\geq 3$.

Let us analyze a second action, again with two derivatives, given by 
\begin{equation*}
\int \Phi \,g_{ad}g_{be}g_{cf}\left[ x^{a},\left[ x^{b},x^{c}\right]
\right] %
\left[ x^{d},\left[ x^{e},x^{f}\right] \right] .
\end{equation*}
It has the following lagrangian 
\begin{equation*}
\mathcal{L}_{2}=\sqrt{\det (g+F)}\,g_{ad}g_{be}g_{cf}\left( \frac{1}{g+F}%
\right) ^{ai}\partial _{i}\left( \frac{1}{g+F}\right) ^{bc}\left( \frac{1}{%
g+F}\right) ^{dj}\partial _{j}\left( \frac{1}{g+F}\right) ^{ef}
\end{equation*}
Again we can expand in powers of $F$. Using the fact that $\left( \frac{1}{%
g+F}\right) ^{ab}=\left( \frac{1}{g-F}\right) ^{ba}$ and following the
computation 
\begin{eqnarray*}
&&\left( \frac{1}{1-F}\right) ^{ia}\left( \frac{1}{1+F}\right) ^{aj}\partial
_{i}\left( F-FF\right) _{bc}\partial _{j}\left( F-FF\right) _{bc} \\
&=&\left( 1+FF\right) _{ij}\partial _{i}\left( F-FF\right) _{ab}\partial
_{j}\left( F-FF\right) _{ab}+\cdots
\end{eqnarray*}
we conclude that 
\begin{equation}
\mathcal{L}_{2}=\mathcal{L}_{1}+o(F^{6}\partial F\partial F).  \label{eq500}
\end{equation}

The purpose of the last two examples if twofold. On one hand, they clearly
show that the invariant actions introduced in the previous section do have
sensible expansions in powers of $F$ and derivatives $\partial $ and are of
the expected general form. In particular, expressions like equation \ref
{eq400} have already appeared in the literature in various settings \cite
{AT1,AT2,T}. On the other hand the above examples, and in particular
equation \ref{eq500}, show that the high $F$ behavior of the action is not
completely determined by the first terms in a power series expansion. It is
nonetheless true that, given a fixed number of derivatives, the number of
possible structures is finite (recall that one cannot introduce
non--derivative brackets $[x^{a},x^{b}]$ in the action). Therefore one needs
to fix in principle a finite number of coefficients in order to fix
completely the high $F$ behavior of the action, given a fixed number of
derivatives.

\section{Conclusions\label{CONCLUSION}}

We have shown that, with the aid of a generalized Poisson bracket, we are
able to construct actions with any number of derivatives which are invariant
under a simplified version \cite{C} of the Seiberg--Witten transformations
described in \cite{SW}. Clearly this is just a first step in a full
classification of invariant actions in the sense described in the
introductory section.

First of all, even within the present setting, one should show that all
invariant actions are linear combinations of the ones described in
section 
\ref{BRACKET}. We have shown that the actions given in terms of $[,]$ are
invariant, but by no means we have claimed that all invariant actions
are of
the form which we have considered. Also, still within the context of the
abelian $U(1)$ theory one should reconsider the above analysis using the
full group of noncommutative gauge transformations. The results in this
paper heavily rely on the geometrically intuitive nature of the simplified
transformation $A\rightarrow \widehat{A}$ which we have considered. In order
to extend the analysis to the more general setting of \cite{SW}, one should
therefore have a better geometrical understanding of the full
Seiberg--Witten transformations. This question is probably intimately related
to an understanding of the invariance problem in the extremely complex
non--abelian case.

\section{Acknowledgments}

I would like to thank A. Connes and M. Douglas for useful discussions and
comments.\pagebreak

\end{document}